# Detection and Classification of Novel Attacks and Anomaly in IoT Network using Rule based Deep Learning Model

Sanjay Chakraborty[1], Saroj Kumar Pandey[2], Saikat Maity[3], Lopamudra Dey[4]

[1]Department of CSE, Techno International New Town, Kolkata, India
[2]Department of Computer Engineering & Applications, GLA University, Mathura, India
[3]Department of CSE, Sister Nivedita University, Kolkata, India
[4]Department of CSE, Meghnad Saha Institute of Technology, Kolkata, India

**Corresponding Author:** Sanjay Chakraborty
**Corresponding Author's Email id:** schakraborty770@gmail.com

*Abstract*
*Attackers are now using sophisticated techniques, like polymorphism, to change the attack pattern for each new attack. Thus, the detection of novel attacks has become the biggest challenge for cyber experts and researchers. Recently, anomaly and hybrid approaches are used for the detection of network attacks. Detecting novel attacks, on the other hand, is a key enabler for a wide range of IoT applications. Novel attacks can easily evade existing signature-based detection methods and are extremely difficult to detect, even going undetected for years. Existing machine learning models have also failed to detect the attack and have a high rate of false positives. In this paper, a rule-based deep neural network technique has been proposed as a framework for addressing the problem of detecting novel attacks. The designed framework significantly improves respective benchmark results, including the CICIDS 2017 dataset. The experimental results show that the proposed model keeps a good balance between attack detection, untruthful positive rates, and untruthful negative rates. For novel attacks, the model has an accuracy of more than 99%. During the automatic interaction between network-devices (IoT), security and privacy are the primary obstacles. Our proposed method can handle these obstacles efficiently and finally identify, and classify the different levels of threats.*

## 1. Introduction

Attackers today make use of complex methods, such as polymorphism, to modify their attack pattern with each new assault they launch. As a result, the identification of previously unknown attacks has emerged as the primary obstacle for cybersecurity professionals and researchers. Recent studies show that signature based, statistics based, anomaly-based and hybrid approaches can be used for the detection of network attacks. Utilizing pre-established attack signatures, signature-based detection systems are put into practice. That is why they cannot detect novel attacks and new variants of known attacks [1]. Machine learning is used in statistics-based detection to gather information from previously identified exploits and establish a baseline for secure system behaviour [31]. Although there is a chance of false positives or negatives with this procedure, its usefulness is restricted. Overall, statistics-based strategies for detecting novel attacks are not very effective. Despite the possibility that anomaly detection systems are useful against novel threats, one of the biggest difficulties is their high false positive rate [2, 3]. Today's hybrid detection methods avoid the shortcomings of the three strategies outlined above while utilizing their various benefits. Typically, hybrid detection systems combine two or three techniques to provide findings that are more reliable [4].

In this article, we have proposed and evaluated a hybrid approach using deep learning and rule-based method to detect and classify known attacks as well as novel attacks with high stages of accuracy, low false positives and false negatives. The model classifies attacks into 3 major classes based on the pattern of network traffic. The classes are "Normal", "Known attack" and "Novel attack". We employ data with three new attack types that are unique to the training CICIDS2017 Dataset to test the proposed model's propensity to recognize novel attacks (also known as attacks that have not been observed previously). Network traffics contain huge volume of data in real life. To handle such large amount of data and classify the different threats and anomalies from them are challenging tasks for typical machine learning algorithms. Therefore, deep learning plays a vital role in terms of efficiency, accuracy and time consumption in this situation. Training time can be less for such huge data compare to ML algorithms. In addition to the expansion of 5G technology, many inexpensive IoT devices can afford to produce considerable amounts of network traffic, which can be exploited for a variety of attacks. It can degrade the performance of the IoT devices and make it vulnerable. To handle such kinds of challenges during IoT traffic, our proposed method can be used.

In summary, contributions of the present study include,
- ✓ A novel end-to-end framework for detection of all types of attacks like known attacks and novel attacks. To the best of our knowledge, this is the first such try using machine learning.
- ✓ The suggested tactic has increased attack detection's precision; reduce the percentage of false positives and false negatives.
- ✓ Our proposed technique requires less time complexity compared to any traditional ML algorithm in case of handling huge amount of network traffic.

The proposed model attacking levels are shown and compared with the existing IDS in Figure 1. Figure 2 and Figure 3 also represent the network architecture of our proposed model that shows the strategy to protect the network from various novel attacks.

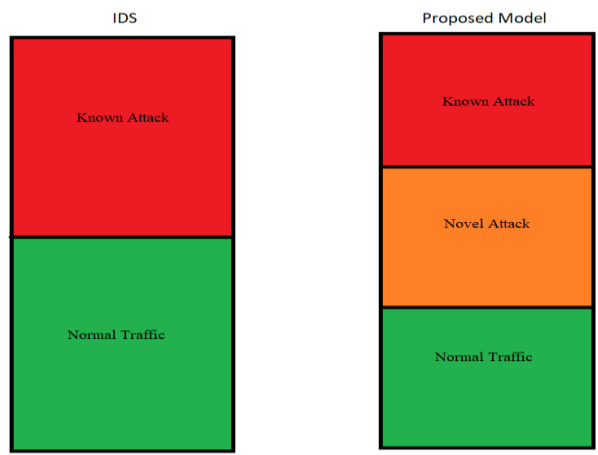

Figure 1: IDS vs Proposed model

The main objective of this work is to introduce and develop a deep learning inspired rule-based network security model that is capable to detect and classify the different categories of network attacks in various types of networks including IoT network. It opens the door of handling several security issues and attacks on different kind of networks.

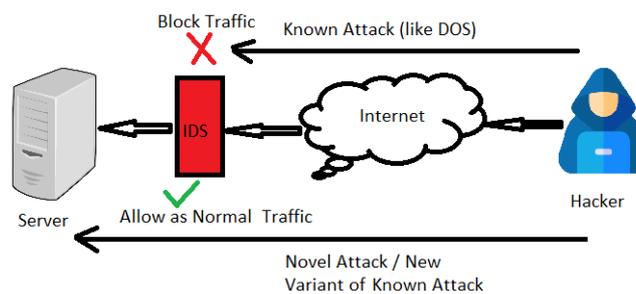

Figure 2: IDS protect network

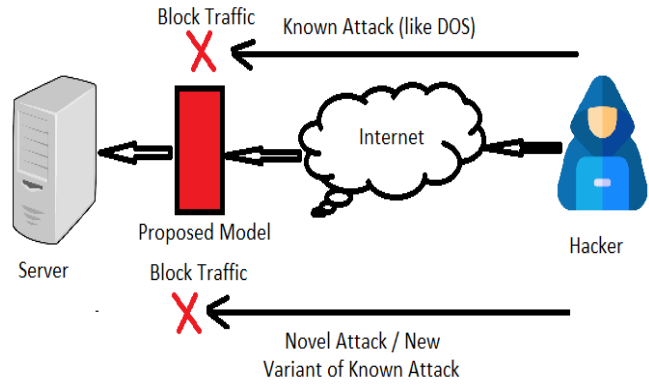

Figure 3: Protection of network by proposed model

This paper is divided into 5 Sections, Section 2 deals with background details and the overall architecture and ruleset of our proposed network model, Section 3 problem statement and type of attacks is defined along with experimented dataset description, Section 4 deals with the experimentation and results analysis. Finally, in Section 5 conclusion is given. At the end of this paper, the readers can find some interesting references.

2.  **Background and Proposed Model**

In previous studies, some of machine learning approaches have been proposed for network attack detection [23]. Some research works use seven different machine learning methods (Naive Bayes, QDA, Random Forest, ID3, AdaBoost, MLP, and K Nearest Neighbours) to detect network anomalies on some popular datasets [5, 6, 28, 29]. Boukhamla use two multi-class classifier MPL model with a different feature set with varying numbers of CICIDS2017 dataset and the model with higher number of features set gives better performance [7]. As a distributed Deep Belief Network (DBN) feature reduction strategy, Marir used different sets. The obtained features are used for a multilayer group SVM. A 60% to 40% split between the training and test datasets in the CICIDS2017 dataset is used [8]. Researchers are also using Neural networks for network attack detection. Pektas suggested merging CNN and LSTM in a deep learning architecture to enhance the performance of the attack detection [35, 36]. The model use CICIDS2017 dataset for testing [9]. Watson uses the Convolutional Neural Network (CNN) classifier and Multi-Layer Perceptron (MLP) classifier with specified packet header features of CICIDS2017 packet capture file [10]. To accomplish attack detection and analysis utilizing deep learning nets and association rule mining, Thilina introduces a unique framework [11]. Zhu uses a CNN model for attack detection and identification. In comparison to conventional machine learning algorithms, this model performs better [12]. To identify port-scan assaults, Aksu suggested a deep learning model and compared the outcomes with the SVM. Deep learning has a total of 30 epochs, a RELU activation function, and 7 hidden layers. The CICIDS 2017 dataset is utilized to train the model. The deep learning model's accuracy rate is 97%, compared to the SVM model's accuracy rate of 67% [13, 14]. Now-a-days several technologies have been developed to handle harmful malware attacks in cloud [22]. In the paper [24], logistic regression and neural network classifiers are mainly used for detecting and preventing threats and anomalies in smart IoT devices. In the paper [25], besides the naïve bayes (NB) and support vector machine (SVM), a unique deep learning algorithm called long-short term memory (LSTM) is considered for anomaly or redundancy detection and modifying attacks in IoT framework [32]. In the paper [26], another stacked-deep polynomial learning based intrusion detection framework is designed and introduced to detect threats in IoT environment. This approach [26] is inspired by the spider-monkey optimization (SMO) technique to choose the optimal features in the dataset and improve the accuracy for the detection of the anomaly. Similarly, a long short term memory (LSTM), BiLSTM, and Gated Recurrent Unit (GRU) techniques are used to introduced an anomaly detection system in IoT networks [27]. Initially, convolution neural network (CNN) is used for analyzing the input features and then the above three deep learning techniques are applied for binary classification of anomaly detection [27]. A very new transfer learning auto-encoder model is introduced to detect noisy DDoS malware attacks such as Mirai and Bashlite in IoT devices [30]. However, both deep learning and machine learning methods have some significant impact on anomaly detection and classification in different networks including IoT [33, 34, 35, 36].

Motivated by the above works, we have developed a rule based deep learning classifier model for similar kind of threats or anomaly detection in networks including IoT environment. The proposed approach classifies among the normal, novel and known attacks based on the historical data of network traffics with high accuracy rates.

According to the Figure 4, the proposed rule-based framework has two-stage. In the first stage, the deep to learn, a learning model is applied attack patterns from the training dataset to find the probability distribution for each attack class of the training dataset. In the second stage, a rule-based model is used to classify the input sample into 3 major classes "Normal", "Known attack" and "Novel attack" based on the probability distribution of the new input sample calculated in the first stage.

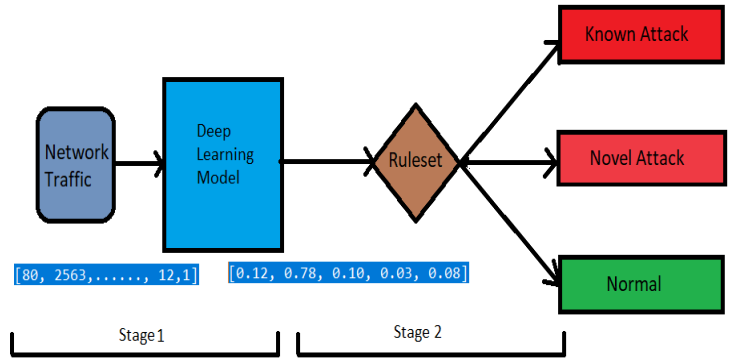

Figure 4: Proposed model architecture

**Ruleset:**

1) Network traffic = "Normal traffic".
    a. Rule: P(Normal) > =.80

2) Network traffic = "Known Attack".
    a. Rule: P (AC 1) / P (AC 2) /. . . . . ./ P (AC N-1) / P (AC N) >= .80
    b. Here N is the total number of attack categories in the training dataset which is 5 for this dataset.
    c. AC mean Attack category and P is the probability of the attack category calculated by the deep learning method.

3) Network traffic = "Novel Attack".
    a. Rule: (P(Normal) < .80) && (P (AC 1) < .80 &&     . . . . . .&& P (AC N) < .80)
    b. In case of a novel attack, the network traffic will be sent to a cyber forensic team for further analysis of the traffic and once the traffic is classified, the model will be trained with the new attack class.

**Procedure:**
**Input:** CICIDS2017 dataset for traffic data.
**Output:** Categorizing three kinds of attacks (normal, novel and known).
Begin
  1. Pre-process the input dataset. Then, split the dataset into training and testing part (70:30).
  2. Apply the proposed rule based deep learning model with a specified architecture on the training dataset.
  3. Measure the performance of the training model through different parameters.
  4. Now, apply the trained deep learning model on the unknown testing data and find the classes of different categories of attacks with the accuracy, recall and FPR measure.
End

### 3. Dataset Description
Due to a paucity of trustworthy datasets, machine learning systems struggle to produce accurate and consistent performance ratings. Some available attack datasets are KDD CUP 1992 datasets, DARPA 1998/1991, BSM98,

BSM99, KDD99, NSDL-KDD, ISCX2012, UNSW-NB15. The majority of the readily accessible datasets are stale and unreliable. Some of them struggle with low traffic volume and diversity, and some of them also lack feature sets and metadata [15]. CICIDS2017 dataset is released by the Canadian Institute for Network security (CIC) in 2017, where it contains benign/normal traffic data and up-to-data common attacks which resemble the true real-world data. As shown in Table 1, CICIDS2017 dataset has the most common attack categories Dos, DDoS, Patator/Brute force, Web-based, Heart-bleed, Infiltration, Bot and Portscan with a total of 14 types of attacks DoSHulk attacks, DoS-Slow HTTPTest attack, DoS-GoldenEye attack, DoS-Slowloris attack, DDoS LOIT, Botnet, FTP-Brute Force, SSH-Brute Force, Brute Force-Web, Web Attack – XSS, Web Attack – SQL Injection, Infiltration and Heartbleed are covered in the dataset. All the data are fully labelled with 78 features (like destination ports, Packet Length, Flow Duration etc.) extracted from the network traffic [16,17].

Table 1: Dataset statistics

| Attack category in CICIDS2017 | |
|---|---|
| Benign/Normal | 3000000 |
| DoS (DoS hulk, DoS slowloris, DoS slowhttptest, DoS goldeneye) | 200000 |
| DDoS | 50000 |
| Patator (FTP, SSH) | 12000 |
| Portscan | 120000 |
| Bot | 1000 |
| Web attack (XSS, Brute force, SQL Injection) | 2000 |
| Infiltration | 36 |
| Heartbleed | 11 |
| Total | 746909 |

The CICIDS2017 dataset is segmented into 3 datasets, a training and testing dataset for the deep learning model, the dataset for probability distribution calculation, sample dataset of novel attack. Training and testing dataset for deep learning model has data of four attack categories (DoS attack, Patator attack, Web-based attack and Portscan attack) and Normal traffic [18,19] shown in figure 5.

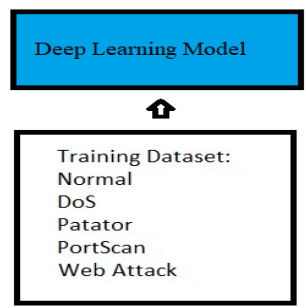

Figure 5: Testing dataset for deep learning model

Dataset for probability distribution calculation has data of four attack categories (DoS attack, Patator attack, Web-based attack and Portscan attack) and Normal traffic. The dataset has unique data [20] shown in figure 6 & figure 7.

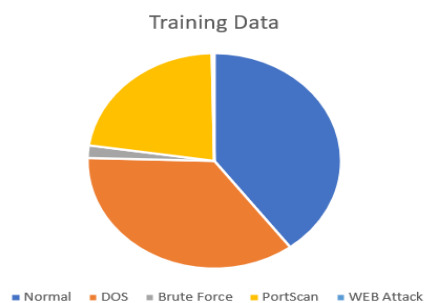

Figure 6: Attack categories in training dataset

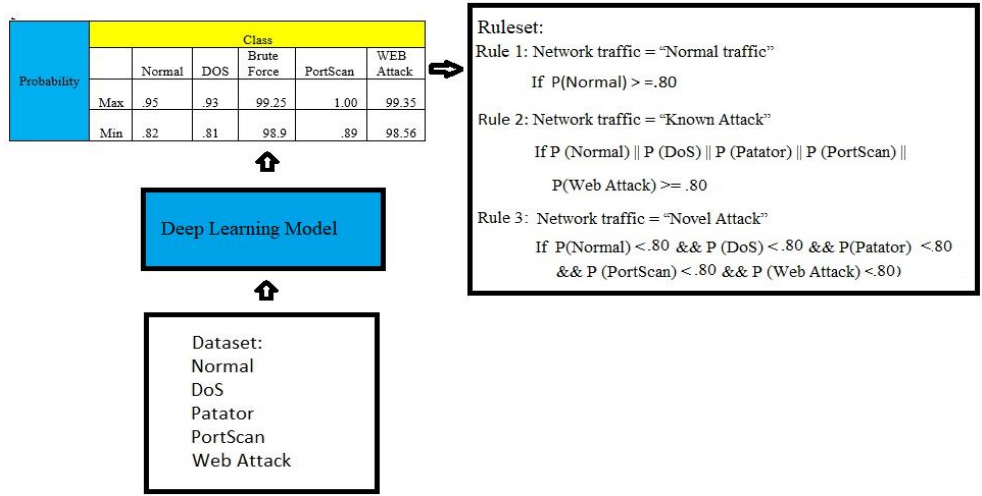

Figure 7: Proposed Attack Model for Attack identification

Sample dataset of novel attack has data of four attack categories (DDoS attack, Bot attack, Heartbleed attack, Infiltration attack). Figure 8 shows the different categories of attacks [21].

## 4. Result Analysis

The proposed model is written in python programming language and python Keras library with TensorFlow backend is used for the deep learning model. All our evaluations are performed on a Windows machine with a quad-core 1.60 GHZ processor, 8 GB RAM. The CICIDS2017 dataset is used for both training and testing of the deep learning model. A brief description of CICIDS2017 dataset is summarized in Table 1. Some records in the collection have values of NaN and infinite. The dataset is cleaned up by removing any entries with NaN values and endless values. Some attributes in the dataset that are derived from the network traffic have unusually wide ranges between their minimum and maximum values. The feature values between 0 and 1 are linearly normalized using a min-max scaling approach to reduce the impact of these outliers.

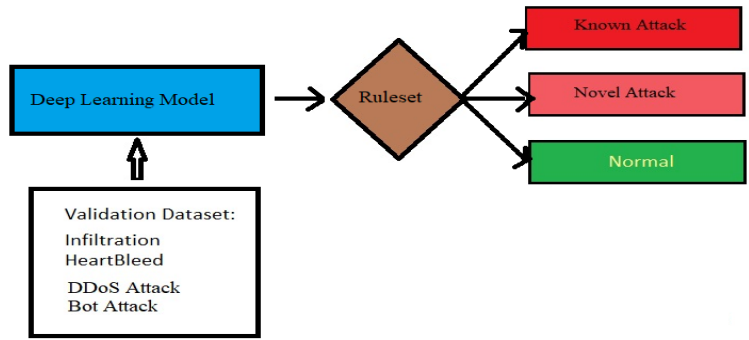

Figure 8. Different categories of attacks classified through ruleset

As shown in Table 2, the output variable of the dataset is a categorical variable, containing 4 different attack labels and 1 normal label. We applied the One-Hot Encoding method to the output variable to transform categorical values into vectors where only one element is non-zero, or hot.

Therefore, the total number of input dimensions is 78 and the output dimension is 5 (4 attack categories and 1 normal). The training dataset of the deep learning model is split into is train and test dataset in the ratio of 70:30. The cross-validation and training datasets are divided from the training dataset in an 80:20 ratio. In addition, figure 8 displays the distribution of attack data in the training dataset. An input layer, several hidden layers, and an output layer make up the deep neural network. 78 neurons make up the input layer, matching the number of input features. Seven hidden layers and the activation function are both part of the deep learning architecture. In the proposed architecture, we have used ReLU activation function in the hidden layer. In comparison to sigmoid and tanh functions, the convergence is quicker. This is so because one linear component of the ReLU function's derivative (slope) is fixed, while the other linear component's derivative is zero. As a result, the ReLU function speeds up the learning process significantly. The output layer and Softmax() function both have 5 neurons, matching the number of attack types in the training dataset. In multi-class issues, Softmax assigns decimal probability to each output class. Because it transforms the scores into a normalised probability distribution and evenly divides the probability among each output node so that the total is 1, Softmax is incredibly helpful. Softmax's output can be seen or utilized as an input by other systems. It is customary to add a Softmax() function as the neural network's final layer because of this. Cross-entropy is employed as the cost function in the deep learning model, while Adam is used as the optimizer with a learning rate of 0.0001. Table 8 represents a detailed description of the used deep learning architecture in tabular form.

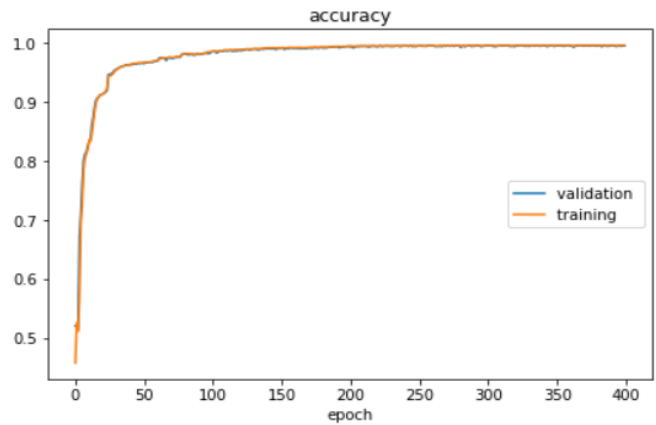

Figure 9. Accuracy vs. Epochs

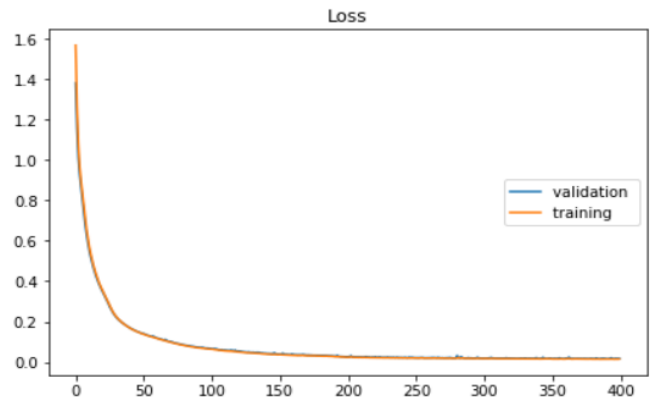

Figure 10. Loss change rate

Figure 9 & figure 10 shows that after the epoch approached 150 times, the model's performance tended to be steady. When the epoch is 400 times, the loss value is 0.0169 and the accuracy value is 0.9958. Table 1 shows the required maximum and minimum probability values required for classifying network traffic as "attack" or "normal". As shown in Table 2, the required maximum and minimum probability values for classifying network traffic as "Normal" are .95 and .82. As shown in Table 3, the required maximum and minimum probability values for classifying network traffic as "DoS attack" are .93 and .81.

|  | precision | recall | f1-score | support |
|---|---|---|---|---|
| Normal | 1.00 | 0.99 | 0.99 | 54479 |
| Dos | 0.99 | 1.00 | 1.00 | 48700 |
| Patator | 0.93 | 0.99 | 0.96 | 2583 |
| Portscan | 1.00 | 1.00 | 1.00 | 30457 |
| webattack | 0.99 | 0.58 | 0.73 | 426 |

Figure 11. Classification Report

Table 2: Probability value range for different attack class

| Probability | | Class | | | | |
|---|---|---|---|---|---|---|
| | | Normal | DOS | Brute Force | PortScan | WEB Attack |
| | Max | 1.00 | 1.00 | .99 | 1.00 | 99 |
| | Min | .93 | .86 | 96.00 | .91 | .93 |

Table 3: Probability values for normal traffic

| Probability | Normal | DOS | Brute Force | Port Scan | WEB Attack |
|---|---|---|---|---|---|
| Max | 1 | 0 | 0 | 0 | 0 |
| Min | 0.93 | 0 | 0.06 | 0 | 0.01 |

Table 4: Probability values for DoS attack

| Probability | Normal | DOS | Brute Force | PortScan | WEB Attack |
|---|---|---|---|---|---|
| Max | 0 | 1 | 0 | 0 | 0 |
| Min | 0.07 | 0.86 | 0.06 | 0.00 | 0.01 |

Table 5: Probability values for PortScan attack

| Probability | Normal | DOS | Brute Force | PortScan | WEB Attack |
|---|---|---|---|---|---|
| Max | 0 | 0 | 0 | 1 | 0 |
| Min | 0 | 0.08 | 0.01 | 0.91 | 0 |

Table 6: Probability values for Patator attack

| Probability | Normal | DoS | Patator | Portscan | Web Attack |
|---|---|---|---|---|---|
| Max | 0 | 0 | 0.99 | 0 | 0.01 |
| Min | 0.01 | 0 | 0.96 | 0 | 0.03 |

As shown in Table 4, the required maximum and minimum probability values for classifying network traffic as a "PortScan attack" are 1.00 and .892. Recall, false negative ratio, and false positive ratio (FPR) are three crucial metrics to evaluate an attack detection system's effectiveness shown in table 5 & table 6. For each novel assault, the suggested model's accuracy is measured in terms of recall, false-negative rate, and false-positive rate shown in fig 11. Recall (R) measures how well the model can identify attacks. It is shown in equation (1).

$$R = TP / (TP+ FN) \ldots\ldots\ldots\ldots (1)$$

When an activity is labelled as an attack by the IDS but is actually just permissible behaviour, this is known as a false positive state (shown in equation (2)). A false alarm is a false positive.

$$FPR = FP / (TP+ FP) \ldots\ldots\ldots\ldots (2)$$

The most serious and hazardous state is a false negative. When an activity is actually an assault, the IDS may mistakenly classify it as permissible behaviour. In other words, a false negative occurs when the IDS misses an attack.

Table 7. Recall & FPR

| Attack Class | Recall | FPR |
|---|---|---|
| DDoS attack | 0.9934 | 0.0221 |
| Bot attack | 0.9928 | 0.0042 |
| Heartbleed attack | 0.9910 | 0.0275 |
| Infiltration attack | 0.9955 | 0.0023 |

Table 8. Proposed Deep Learning Model Architecture

| Layers | No. of Neurons/layers | Activation Function | Learning Rate | Optimizer | No. of epochs |
|---|---|---|---|---|---|
| Input Layer | 1 input layer contains 78 neurons | - | - | - | - |
| Hidden Layers | 7 | ReLU | 0.0001 | Cross-entropy and ADAM | 30 |
| Output Layer | 1 (Five dimensions) | Softmax | - | - | - |

The above result shows a low false positive (FP) rate of less than 0.025 and a high true positive (TP) rate of more than 98.5%. The suggested ruled-based architecture maintains a great trade-off between attack detection and false positive rates, as illustrated in Table 7's performance evaluation findings and conclusion. The result shows that deep learning models could be used to detect existing as well as new network attacks. In comparison to existing traditional machine learning techniques, our model improves network attack detection accuracy while lowering false positive and true positive rates. The system's overall accuracy is 99.5%, while the recall rate for the four categories is 99.9%. The ruled-based model performs well to classify the Infiltration and Heartbleed novel attacks. A higher attack detection rate of the model can be achieved by training the model with a dataset that has a large number of diversified attack classes and data.

In the Table 9, we have shown a comparison with some existing approaches on our experimented dataset (CICIDS 2017). In this comparison, it is clearly seen that our proposed approach provides a stronger security mechanism due to the inclusion of the rule based approaches and due to the impact of deep learning algorithms, it provides much better accuracy and speedup compared to all the existing ML approaches.

Table 9. Comparison in terms of accuracy and other properties on CICIDS2017 dataset

| Approaches | Accuracy | Parameters |
|---|---|---|
| Transfer Auto-encoder Neural Network [30] | 99.98% | transfer learning takes 47.31% and 58.27% less time compared to traditional deep learning |
| ML based Techniques ([1, 2, 13, 24, 28, 29]) | ~ 90% (avg) | More time required to handle huge network traffic. |
| SMO inspired DL [29] | 99.02% | Time complexity is high due to SMO |
| Proposed Method | **99.5%** | - Overall time is less compared to and ML based techniques.<br>- Rule based approach makes the security clauses stronger. |

This proposed approach has several usage in real-life applications. Applications of such kind of anomaly detection include finding possible risks or medical issues in health data, detecting faults in manufacturing, detecting intrusions into computer networks, monitoring sensor readings in aeroplanes, and predictive maintenance. Monitoring any data source, including user logs, devices, networks, and servers, is possible thanks to anomaly detection. Quickly recognise unknown security dangers as well as zero-day attacks. Discover anomalous activities across data sources that are missed by conventional security techniques.

## 5. CONCLUSIONS

Detection and classification of various network attacks plays a vital role for efficient and effective network traffic communication. This work focuses on the development of deep learning and rule-based network security model that helps to find the network traffic patterns to detect and categorize the different kinds of network attacks. The proposed model is evaluated and cross validated for accuracy in terms of the percentage of false positives and false negatives. Our proposed model achieves a very attractive attack detection rate on the benchmark experiment dataset. The proposed system has a 99.5% average accuracy rate, and a 99.9% recall rate for the 4 categories of attacks. A low FP rate (< 0.025) and a high TP rate (> 98.5%) are also displayed in the results. The CICIDS2017 dataset alone does not offer enough reference points for this suggested model to develop a comprehensive picture in general. While detecting anomalies, this model can exhibit varied behaviours on real-time traffic data.

In the future, we plan to test the proposed approach over a set of different classes of network traffic, and different types of novel attacks and evaluate in detail the attack prediction capability of the model. For anomaly detection on the real-time traffic, we plan to convert the raw-traffic files into feature vectors through IPFIX framework in future as it can configure unique properties with ease. This would make for an intriguing foundation for creating a future feature list that is more comprehensive. Besides that, RNN/LSTM use can be applied to handle the anomalies on time series IoT traffic data.

**Conflicts of Interest**: The authors declare no conflicts of interest.

**Funding Statement**: All authors declare that there is no funding available for this article.